\documentclass[preprint]{aastex}
\usepackage{psfig}

\newcommand{\ha}{H$\alpha$~}

\shorttitle{Andromeda IV}
\shortauthors{Ferguson et al.}

\begin{document}

\title{On the Nature of Andromeda~IV{\footnote{Based on observations
with the NASA/ESA Hubble Space Telescope, obtained at the Space
Telescope Science Institute, which is operated by the Association of
Universities for Research in Astronomy, Inc. under NASA contract No.
NAS5-26555.}}~$^,${\footnote{Based on observations made with the William Herschel
and Isaac Newton telescopes operated on the island of La Palma by the
Isaac Newton Group in the Spanish Observatorio del Roque de los Muchachos
of the Instituto de Astrofisica de Canarias.}}}



\author{Annette M.~N. Ferguson}
\affil{Institute of Astronomy, University of Cambridge, Madingley Road,\\
Cambridge, UK CB3 0HA}
\email{ferguson@ast.cam.ac.uk}

\author{J.~S. Gallagher}
\affil{Dept. of Astronomy, University of Wisconsin at Madison, 475 North 
Charter Street, Madison, WI 53706}
\email{jsg@astro.wisc.edu}

\author{Rosemary F.~G. Wyse}
\affil{Dept. of Physics \& Astronomy, Johns Hopkins University, 
3400 North Charles Street, Baltimore, MD 21218}
\email{wyse@pha.jhu.edu}

\begin{abstract}

Lying at a projected distance of 40\arcmin\ or 9~kpc from the centre of
M31, Andromeda~IV is an enigmatic object first discovered during van
den Bergh's search for dwarf spheroidal companions to M31.  Being
bluer, more compact and higher surface brightness than other known
dwarf spheroidals,  it has been suggested that And~IV is either a
relatively old $`$star cloud' in the outer disk of M31 or a background
dwarf galaxy. We present deep HST WFPC2 observations of And~IV and the
surrounding field which, along with ground-based long-slit spectroscopy
and H$\alpha$ imagery, are used to decipher the true nature of this
puzzling object.  We find compelling evidence that And~IV is a
background galaxy seen through the disk of M31.   The moderate surface
brightness ($\bar{\mu_V}\sim$24), very blue colour
(V$-$I$\lesssim$~0.6), low current star formation rate
($\sim$0.001~M$_{\odot}$~yr$^{-1}$) and low metallicity
($\sim$10\%~Z$_{\odot}$) reported here are consistent with And~IV being
a small dwarf irregular galaxy, perhaps similar to Local Group dwarfs
such as IC~1613 and Sextans~A.   Although the distance to And~IV is
not  tightly constrained with the current dataset, various arguments
suggest it lies in the range 5$\lesssim$D$\lesssim$8~Mpc, placing it
well outside the confines of the Local Group.  It may be associated
with a loose group of galaxies, containing major members UGC~64,
IC~1727 and NGC~784.  We report an updated position and radial velocity
for And~IV.

\end{abstract}



\section{Introduction}

Andromeda~IV is an enigmatic object first discovered by \citet{vdb72}
during his photographic search for dwarf spheroidal companions to
M31.   On deep optical plates, it appears as a faint smudge lying at a
projected distance of 40$'$ or $\sim$9~kpc (assuming D$_{M31}=$784~kpc,
\cite{holl98,stan98}) to the south-west of  M31 (see Figure 1).
Noted immediately as being more compact, bluer and of higher surface
brightness than the other dwarf spheroidals he identified,  van den
Bergh suggested And~IV was either a relatively old  $`$star cloud' in
the outer disk of M31\footnote{And~IV actually appears listed as the
open cluster C188 in the Atlas of the Andromeda Galaxy \citep{hod81}.}
or a background dwarf galaxy.    If in the disk of M31, And~IV would
have a size of $\sim$200~pc and lie at a deprojected distance from
the center of 25~kpc ($\sim$5 disk scalelengths, beyond the optical
edge of the stellar disk), thus representing an example of a large diffuse star
cluster which has formed in the outer regions of a galaxy where there
is very little star formation at the present epoch.  Such an object
could potentially yield important insight into the nature of the star
formation process under the extreme physical conditions of  low gas
surface density, high gas fraction and long dynamical timescales.  On
the other hand, And~IV could be yet another example of a previously
uncatalogued dwarf galaxy lurking in our local environs, perhaps even
bound to M31 (eg.  \citet{arm98,arm99}).  Determining the distance,
constituent stellar populations and evolutionary state of this puzzling
object is therefore of obvious importance.
 
Ground-based study of And~IV has been severely hampered due to the
presence of a  bright (V$\sim$10) foreground star lying within 30$''$
(see Figure 1) and by the combination of faintness and
crowding of M31 stars along the line of sight.  \citet{j93} used CFHT
imagery to construct a colour-magnitude-diagram (CMD) to V$\sim$23 of
the And~IV region, which he interpreted as representing a young
population of stars with a narrow age range, an $`$unusually large'
open cluster in M31.  Jones's data, however, did not allow a proper
statistical subtraction of M31 field stars  to be carried out and thus
the possibility remains of a significant foreground M31 contamination
in his And~IV CMD.  The  reported HI detection of And~IV is also rather
uncertain.   In their catalogue of extragalactic HI observations,
\citet{hr89} list an HI radial velocity for And~IV of $-$375~km/s.
This measurement can be traced back to the early work of
 \citet{emer74}, who used the Cambridge Half-Mile telescope to map the
disk of M31 with a spatial resolution of
1.5\arcmin~$\times$~2.2\arcmin.  \citet{emer74} notes that And~IV
appears projected near a faint outer HI arm of M31 (see also
\citet{unw80}) and that the measured velocity of the HI along this arm
($-$375~km/s ) is consistent with that expected at that location in the
disk based on the major axis rotation curve.  The case for the
association of this gas with And~IV, as opposed to merely M31's disk,
would appear to have no stronger foundation than this, and therefore
must be regarded as weak.

Detailed study of the structure and stellar populations of And~IV
requires the high resolution imaging capability of {\it Hubble Space
Telescope}.  We were awarded 5 orbits of Cycle 6 HST/WFPC2 time to
observe And~IV and the surrounding M31 field. We present these data
here, along with supporting ground based observations (H$\alpha$
imagery and optical long-slit spectroscopy). This new dataset clearly
resolves the true nature of And~IV, for the first time, as a small
background dwarf irregular galaxy unassociated with M31. A companion
paper will present an analysis of the field stellar populations in the
outskirts of M31, derived from the same HST/WFPC2 dataset
(Ferguson, et al 2000, in preparation).

\section{Observations and Reductions}

\subsection{HST/WFPC2 Observations}

HST/WFPC2 images were taken of And~IV and the surrounding M31 field
over five orbits on October 31, 1998 (GO \#6734).  The 3 WF cameras
provide an L-shaped FOV of 150\arcsec\ by 150\arcsec\ with 0.1\arcsec\
pixels, while the PC provides a square 34\arcsec\ by 34\arcsec\ FOV
with 0.046\arcsec\ pixels.   Total exposure times amounted to 6000s and
6100s in the F555W (WFPC2 broadband V) and F814W (WFPC2 broadband  I)
filters respectively. The proximity of a very bright star ($\sim$10th
mag) to And~IV imposed stringent roll angle constraints on our
observations in order to avoid excessive scattered light and
bright-object artifacts in our primary region of interest.  The optimal
orientation of the camera placed And~IV largely on the WF3 chip and the
bright star on the edge of WF4, a strategy which rendered the WF4 chip
essentially useless but kept the region around And~IV free of
diffraction spikes and other artifacts.  The WF3-FIX aperture  was
centered on $\alpha_{2000}=$00$^h$42$^m$30.1$^s$ and
$\delta_{2000}=+$40$^{\circ}$34\arcmin32.7\arcsec.  Both the WF2 and
the PC imaged the surrounding field populations of M31.

Images were processed through the standard STScI pipeline.   The frames
were split into images of each individual CCD and the vignetted regions
of the chips set to zero.   The data were taken undithered and
measurements of several bright stars confirmed no shifts were required
to align the individual frames before combining.  Images taken through
a given filter were combined using the cosmic-ray rejection algorithm
CRREJ in IRAF \footnote{IRAF is distributed by the National Optical
Astronomy Observatories, which are operated by the Association of
Universities for Research in Astronomy, Inc., under cooperative
agreement with the National Science Foundation.}.  Corrections were
applied for warm pixels using pixel lists provided by STScI that were
generated closest to the date of the observations.  Figure \ref{wfima}
shows a colour representation of the combined F555W and F814W WF3
images; AndIV is clearly visible as a diffuse blue concentration
towards the lower left of the chip.

\subsection{HST/WFPC2 Photometry}  

Photometry was performed using the IRAF implementation of the
crowded-field photometry package DAOPHOT/ALLSTAR \citep{stet87}.  Stars
were detected on each chip using a DAOFIND threshold of
5$\sigma_{bkgd}$, a value found to maximise the number of real stars
detected while maintaining a low level of spurious sources.  Given that
our field is relatively crowded ($\sim$7000$-$8000 stars per WF chip),
there were few truly isolated bright stars suitable for characterizing
the point-spread function (PSF).  Our approach was instead to select a
set of $\sim$30 of the brightest stars on each chip, subtract out all
the other stars using a first-guess PSF and then use the remaining
$`$clean' stars to construct a more refined PSF model.  This was done
for each filter/chip combination.  The PSF models were based on Moffat
functions with $\beta=$1.5 and a look-up table of residuals, and these
shapes were held constant across each chip.  ALLSTAR photometry was
carried out using the parameter values recommended by \citet{ck95} to
maximize the performance.  We also experimented with using Tiny Tim
PSFs for the photometry instead of PSFs built from the images
themselves, and found that the results were generally very similar.

Aperture corrections were derived in the same manner to that used to
construct PSF models, namely by isolating the brightest stars on each
chip and subtracting out the rest.  Total magnitudes were measured for
the remaining stars using an aperture of radius
0.5\arcsec\ \citep{h95b} and aperture corrections were defined in the
sense $m_{PSF}-m_{0.5\arcsec}$.  As the focus and PSF shape is a
function of position on the WFPC2, aperture corrections were allowed to
vary linearly across each chip.  Corrections were also applied for
geometric distortion and charge transfer efficiency using the
recommendations of \citet{h95a} and \citet{whit99} respectively.
Before transforming  the instrumental magnitudes to the standard
system, we corrected our data for the effects of extinction.  We
adopted  a foreground Galactic reddening of E(B$-$V)$=$0.08$\pm$0.02
towards M31\footnote{Note that \citet{sch98} quote the slightly lower
value of E(B$-$V)$=$0.062 for the Galactic reddening towards M31,
derived from the mean dust emission detected by DIRBE in surrounding
annuli.}.  \citep{bh84} and used Table 12(b) of \citet{h95b} to derive
the corresponding extinctions in the HST bandpasses.  As And~IV lies
well beyond the main HI disk of M31, in a region where the mean
N$_{HI}\lesssim$3$\times$10$^{20}$~cm$^{-2}$,
\citep{emer74,unw80,sof81}, the reddening due to M31's disk
 is low for all reasonable gas-to-dust ratios and hence we apply no
correction.  As And~IV appears projected on a faint outer HI arm, there
does exist the possibility of differential reddening across the face of
the object \citep{emer74,unw80}. Unfortunately, we have no way to
estimate the magnitude of this effect but it should be kept in mind.
Transformation to standard Johnson-Cousins V- and I-band  magnitudes
was carried out via the iterative approach described in \citet{h95b},
and adopting the values presented in their Table 7.

Our photometry reaches to V$\sim$27.5, I$\sim$26.5.   Typical
1-$\sigma$ photometric errors for stars at 25th magnitude are
$\sigma_V\sim$0.06 and $\sigma_I\sim$0.08.  Only stars for which
ALLSTAR photometry was deemed high quality ($\chi<$2 and $-$0.2$<${\it
sharpness}$<$0.2) are used in our final analysis. The completeness
level at faint magnitudes is not a serious concern  for the analysis
that we present here,  and we postpone a detailed discussion of this
issue to a  future paper.  Past experience with WFPC2 photometry
leads us to expect $\gtrsim$80\% completeness above V$\sim$26 and
I$\sim$25.5 \citep{dp97,cole99}.  Our analysis in this paper focuses
only on those stars detected in the WF3 chip.

\subsection{Ground-based Follow-Up Observations} 

We also obtained complementary  ground-based  imaging and spectroscopy
of the And~IV region.  Deep narrow-band observations were obtained via
the ING service program in August 1999 using the INT~2.5m telescope
equipped with the Wide Field Camera at the f/3.3 prime focus.  The Wide
Field Camera consists of 4 thinned EEV 4k$\times$2k CCDs, each covering
an area of $\sim$23\arcmin$\times$11\arcmin\ on the sky with
0.33\arcsec\ pixels.  Our images therefore include  a large portion of
the surrounding M31 field. Exposures of 3$\times$800s were taken
through a narrow-band \ha filter ($\lambda_c=$6568{\AA},
$\Delta\lambda=$95{\AA}) and a 300s exposure through broadband Sloan
{\it r$\prime$} for the continuum subtraction.  Conditions were
photometric, but poor seeing prevailed ($\sim$2$-$2.5\arcsec).  The
images were reduced in the standard manner (see for example,
\citet{fer96a}) and observations of the spectrophotometric standard
Feige~110 from the list of \citet{mass88} were used for the photometric
calibration.  The average sensitivity of the \ha continuum-subtracted
image, taken to be 1$\sigma$ of the sky background, is determined to be
$\sim$2.7$\times$10$^{-17}$~erg~s$^{-1}$~cm$^{-2}~/\Box\arcsec$.
Figure \ref{haima} shows portions of both the unsubtracted and the
continuum-subtracted \ha images centered on And~IV, with the WFPC2
field-of-view overlaid on the continuum-subtracted image.  Several
faint emission line sources are clearly visible in the vicinity of
And~IV.

We obtained long-slit optical spectroscopy of several of these objects
in September 1999 using the WHT~4.2m and the ISIS double-beam
spectrograph.   The R300B grating was used with a 4k$\times$2k thinned
EEV CCD in the blue arm to cover the range $\sim$3700{\AA}$-$5300{\AA}
with 3.4{\AA} resolution; in the red arm, we used the R316R grating and
a 1K$\times$1K TEK CCD to cover $\sim$5700{\AA}$-$7200{\AA} with
3{\AA} resolution.  The slit length and width were 3.7\arcmin\ and 1\arcsec\
respectively.  Conditions were excellent  with sub-arcsecond seeing.
Most observations were made at very low airmass to minimize the effects
of differential atmospheric refraction. In addition, the slit was
rotated close to the parallactic angle whenever the airmass exceeded
1.2.  Total exposure times ranged from 1800$-$6300s, depending on the
brightness of the source.  Spectra of CuAr and CuNe lamps were made
during the night to provide a wavelength calibration and observations
of spectrophotometric standards from the list of \citet{mass88} were
made for the flux calibration.

Given the faintness of the emission line sources, special care was
required during the processing of the spectra.  Individual
2-dimensional spectra from a given slit position were aligned to the
nearest integer pixel and the sky removed by fitting a low order
polynomial fit with the IRAF BACKGROUND task. The spectra were
subsequently averaged together using a cosmic-ray rejection algorithm
and apertures ranging from 3-4\arcsec\ were used to extract
one-dimensional spectra of each emission-line object.  Care was taken
to ensure that red and blue spectra for a given  object  corresponded
to the same physical apertures (both size and shape).  A region of
$\sim$20\arcsec\ centered on the brightest continuum emission from
And~IV was also extracted.  The one-dimensional spectra were wavelength
and flux calibrated with residuals of $\sim$0.5{\AA} and 0.02-0.04 mag
respectively.  In Figure \ref{specplts}, we show the calibrated spectra
for the three brightest nebulae (\#3,4,6) in the vicinity of And~IV as
well as for the underlying continuum emission.

\section{Results} 

\subsection{Stellar Populations}

\subsubsection{Resolved Emission}

The (V,V$-$I) and (I,V$-$I) colour-magnitude diagrams (CMDs) of 6388
stars detected in the WF3 chip are shown in Figure \ref{cmds}.  These
CMDs reveal a prominent red giant branch (RGB) with a significant
intrinsic width,  a red clump, a weak blue plume and a possible blue
horizontal branch.  The striking downturn  of the RGB at red colours
(V$-$I$\gtrsim$2) has previously been seen in the outskirts of M31
(eg.  \citet{holl96}) and indicates the presence of a metal-rich
component to the stellar population.  Based on standard star count
models \citep{gil81,wys89}, we expect $\sim$20 stars redder than the
RGB and $\sim$1 star bluer than the RGB between
21$\lesssim$V$\lesssim$27 along this line of sight towards M31.
Galactic field stars are therefore expected to contaminate our field by
a negligible amount ($\sim$0.3\%).

It is difficult to tell from visual inspection of Figure \ref{wfima}
alone what fraction of And~IV has resolved on our deep images.  In
order to isolate the signature of And~IV stars  from those of the M31
disk, we constructed CMDs for stars lying within a box of
40$''\times$60$''$ centered on the brightest emission from And~IV and
for stars lying on the remaining area of WF3, which we will refer to as
the $`$M31 field' (see Figure \ref{compcmds}).   This physical region
was selected as it completely encompasses the area of $`$diffuse'
emission which defines And~IV in Figures 1 and 2.  Taking account 
of the slighly differing areas (0.9:1), we find that the number of
stars detected in each region  of the chip is largely consistent with a
uniform distribution of stars, and indicates only a $\sim$10\%
enhancement in star counts (at least to the limits of our photometry,
V$\sim$27.5) in the neighbourhood of  And~IV.  This is a strong
indication that only a fraction of  And~IV has resolved.
  
Comparison of the morphology of the CMDs in the different areas of the
chip reveals some puzzling differences however.  The boxes in Figure
\ref{compcmds} indicate regions of the CMDs that are populated by stars
in the vicinity of And~IV  but not in the M31 field.   There are, for
example, considerably more faint stars redward of the RGB
(V$-$I$\gtrsim$1.3, V$\gtrsim$25) on the And~IV CMD than on that of the
M31 field (159 stars versus 44).  Likewise, the region in between the
blue plume and the RGB (V$\lesssim$25, 0$\lesssim$V$-$I$\lesssim$0.7)
is also more populated on the And~IV CMD (45 versus 7 stars) as is the
region just above the RGB (V$\lesssim$2.5,
0.8$\lesssim$V$-$I$\lesssim$2.1, 9 versus 1 star).  These overdensities
are significant, especially when account is made for the slightly
differing physical areas that the stars on each CMD are drawn from.

We first investigate whether the stars detected in the vicinity of
And~IV simply have larger photometric errors than those in the
surrounding M31 field, causing them to exhibit broader blue plumes and
red giant branches and scatter to both brighter and fainter
magnitudes.   The increased and more variable sky background in the
And~IV region of the chip could possibly lead to this effect, as could
increased crowding. To test for this, we calculated the mean
photometric errors returned by ALLSTAR in 0.5 magnitude bins for stars
lying on and off the And~IV area.  While the magnitude errors are very
similar over most of the magnitude range, they  start to diverge
towards faint magnitudes.  Still, the effect is small. At V$\sim$27,
photometric errors for stars near And~IV differ by only
$\sim$0.05 mag from stars located elsewhere on WF3.  Thus, while
increased photometric scatter can partially explain some of the faint
stars redward of the RGB, it cannot explain all of them.  Furthermore,
increased photometric uncertainties seem an unlikely explanation for
the excess populations of stars seen in other regions of the CMD.

We then address the issue of whether the $`$excess' populations of
stars seen in the And~IV CMD could be the signature of a distinct,
partially resolved stellar population lying at a significant distance
beyond M31?  Indeed, the stars lying within the marked boxes of the CMD
in Figure \ref{compcmds} account for roughly half the observed stellar
overdensity seen towards And~IV.  These stars could appear offset from
the M31 main sequence and RGB due to internal extinction within either
the galaxy itself, or due to variable small-scale extinction in the
foreground disk of M31.   We constructed RGB and main-sequence
luminosity functions (LFs) in 0.5~mag bins  for stars lying within the
box centered on And~IV and those lying elsewhere on the chip (Figure
\ref{lumfs}). We crudely define RGB stars as those with V$-$I$>$0.6 and
main-sequence stars as those with V$-$I$<$0.6 (this definition will
also allow the inclusion of blue horizontal branch stars into the
$`$main sequence' sample) and normalise the counts in each region for
their slightly different areas.  Comparison of the main-sequence LFs
indicates a genuine excess of blue stars in the vicinity of And~IV with
respect to the M31 field over the entire magnitude range probed.    On
the other hand, inspection of the bottom panel of Figure \ref{lumfs}
reveals no obvious excess of red stars towards And~IV, but does show an
intriguing $`$bump' on the tail of the And~IV RGB LF beyond the red
clump, at I~$\sim$~25$-$25.5.   This feature also appears in the V-band
RGB LF at V~$\sim$~26$-$26.5, but it is \emph{not} present in the M31
field LF.  Quantitative study of this bump is difficult.  In this
magnitude range, incompleteness is an important factor. In fact, it is
probably more of a serious issue in the immediate area of And~IV, where
the background is higher, than it is elsewhere on the chip; the fact
that the bump remains prominent in both passbands suggests it is real.
The colour of the feature, V$-$I~$\sim$~1, is suggestive of a
population of red giants and we are tempted to speculate that the bump
represents a detection of the tip of the red giant branch, or possibly
a slightly more luminous extended asymptotic giant  branch, in a
stellar system located at some distance behind M31.  We will return to
this issue in more detail in Section \ref{disc}.

From the analysis of number counts, CMDs and LFs for stars detected in
our WFPC2 images,  we therefore conclude that there is evidence for
only a small enhancement in stellar density towards And~IV.
Furthermore, a significant fraction of this excess population appears
displaced from the main sequence and red giant branch  of M31 field
stars.    While this may imply that And~IV is not a star cluster lying
in M31's disk,  we are not able to rule out the possibility of a
highly-skewed mass function which would allow And~IV to still be
associated with M31 but not to resolve to the same extent.  We
therefore turn our attention to other aspects of the WFPC2 data, as
well as to other data, in order to derive additional constraints on the
nature of the object.

\subsubsection{Unresolved Emission}

One of the outputs of the ALLSTAR PSF-fitting photometry package is a
residual image where all detected and photometered stars have been fit
with the adopted PSF and subtracted out.  Figure \ref{resima} shows the
residual WF3 F555W image, which has been smoothed with a box of 10
pixels ($\sim$ 1$''$) to remove the residuals of subtracted stars and
to increase the signal-to-noise of the faintest emission.  A large
fraction of unresolved light very clearly remains around And~IV; the
light distribution appears to be somewhat centrally-concentrated with a regular
structure.   By comparing the emission remaining in the residual image
with that in the original image, we find that only 40\% of the F555W
light in the And~IV region of WF3 has resolved, as compared with
70-80\% of the light elsewhere on the chip.  For the F814W image, these
numbers are 50\% and 80$-$90\% respectively. This confirms our earlier
conclusion that only a small fraction of And~IV has resolved compared
to the M31 field.  From the smoothed image, we are able to define a
more accurate centre for And~IV, which we report here as
$\alpha_{2000}=$00$^h$42$^m$32.3$^s$ and
$\delta_{2000}=+$40$^{\circ}$34\arcmin18.7\arcsec.

The unresolved emission of And~IV is of moderately  low surface
brightness and very blue colour.  Using the residual F555W and F814W
images, we measure $\bar{\mu_{V}}\simeq$ 24.0, $\bar{\mu_{I}}\simeq$
23.4 and $\overline{V-I}\simeq$ 0.64 within a 30$''\times$34$''$ box
centered on And~IV.  We have constructed crude surface brightness
profiles for And~IV using the 10$\times$10 pixel boxcar-smoothed
images.  Elliptical aperture photometry was carried out using a fixed
position angle and ellipticity, both of which were determined by eye to
best match the poorly-defined And~IV isophotes.  The upper panel of
Figure \ref{sbprofs} shows the V and I-band surface brightness profiles
derived out to a radius of 20\arcsec\ (the  extent of the bright residual
emission on our WFPC2 image) from the center of And~IV. The
profiles are observed to decline slowly and smoothly with increasing
galactocentric radius.  A model of an exponential disk with
$\mu_{0}=$23.3 and $\alpha^{-1}=$11\arcsec\ is overplotted; such a
light profile appears to provide a good match to the V-band light of
And~IV.  The lower panel of Figure \ref{sbprofs} shows the V$-$I radial
colour gradient and indicates a gradual trend of bluer colours towards
larger radii, although the S/N ratio becomes very low in these
regions.

Surface brightness profiles of open clusters are well-fit by King
profiles (eg. \cite{mat84}), reflecting their tidal limitations.  On
the other hand,  small dwarf irregular galaxies are typically
characterised by approximately exponential profiles and small colour
gradients (eg. \citet{bre99}).  It thus seems that the exponential
profile we find here is further evidence against And~IV being an open
star cluster in M31; instead, it may support the idea that it is a
background dwarf galaxy.

\subsection{Ionized Gas}  

Our deep \ha images reveal eight compact emission-line sources in the
general vicinity of And~IV, five of which lie within 1\arcmin\ of our new
adopted centre (Figure \ref{haima}).  This clustering of emission-line
sources represents a significant overdensity compared to the rest of
the field contained in our wide-field images, and suggests a connection
between at least some of the sources and the faint diffuse continuum
emission which defines And~IV.  Table \ref{emsrcs} lists the positions
and fluxes of the \ha sources, their projected distances from the
adopted centre of And~IV and indicates whether the source appears to be
resolved on our images and whether it has a continuum counterpart.  The
\ha\ fluxes have been measured within a circular  aperture of radius
5\arcsec, and have been corrected for both Galactic extinction using
\citet{bh84} and [NII] contamination using the mean [NII]/H$\alpha$
ratio from our spectra (see below). The \ha fluxes range from
1$-$9$\times$10$^{-15}$~erg~s$^{-1}$~cm$^{-2}$; if at the distance of
M31, these sources would have only modest \ha luminosities of
9$-$68$\times$10$^{34}$~erg~s$^{-1}$ which, for reference, 
are 10$-$100 times fainter than that of the Orion nebula
\citep{kenn84}.  Assuming they are ionization-bounded, such objects
would be powered by Lyman continuum luminosities, Q$_0$, of
$\sim$7$-$50$\times$10$^{46}$ photons~s$^{-1}$ \citep{lh95}, which
could possibly be provided by either early B
stars\footnote{Unfortunately, predictions of ionizing fluxes for stars
of spectral types later than $\simeq$B0 do not yet exist, and the few
direct observations of the Lyman continuum spectral region in B stars
have produced very ambiguous results (see discussion in
\citet{sdk97}). The latest spectral type considered by \citet{sdk97} is
a B0.5~V star, which they predict produces an ionizing flux of
log~Q$_o$(s$^{-1})~\sim$~47.8.} or the luminous central stars of
planetary nebulae \citep{sdk97,vacca96,men97}.  A concentration of PNe
within such a small area of the sky seems rather unlikely however.

If we assume ionization by a single massive star in M31, the \emph
{most} luminous star we could expect to find associated with each
nebula would be of type B0.5V with a magnitude of V$\simeq$20 at that
distance \citep{vacca96, sdk97}.  We would not expect to find stars
much fainter than this, due to the rapid drop-off in ionizing flux as a
function of spectral type. Four of the emission-line sources around
And~IV (\#3,4,5,6) lie in the area of the sky covered by our WFPC2
image and so we were able to search our images for luminous and/or very
blue stars and compact clusters in the vicinity of each nebula. Objects
\#3, 4 and 6 are easily recognizable on Figure \ref{wfima} as being
very blue, and in the case of \#3 and 4, also extended.  As the F555W
filter contains the emission lines [OIII]$\lambda\lambda$4959,5007 and
H$\beta$, with even a small transmission at H$\alpha$, the nebular
morphologies seen at these positions is not surprising.  We base our
search on the results of aperture photometry, as opposed to PSF-fitting
photometry, since the possibility exists that the ionizing sources may
be clusters that would not be well-fit by the PSF model; indeed such
objects have been deliberately excluded from our final stellar
photometry lists (see Section 2.2).  The brightest blue sources
identified lying within 2\arcsec\ of the position of each nebula are
found to have magnitudes V$\simeq$22.4, 22.6 (Object \#3), 23.0 (Object
\#4) and 24.0 (Object \#6), all with V$-$I~$\lesssim~-$0.1.  At the
distance of M31, these sources would have spectral types ranging from
mid to late-B, and as such, would have difficulty in  producing
the required ionizing fluxes.  There are no very blue stars lying
within 2\arcsec\ of Object \#5 however there is a luminous red star
with V$\simeq$22.4 and V$-$I$\simeq$1.6.  Interestingly, this object
lies in one of the areas of $`$excess' stars identified in Figure
\ref{compcmds}.   Given the considerable stellar density in our WFPC2
field, the likely uncertainties in our absolute positions of the
emission line sources and the possibility of differential reddening
across the face of And~IV, it is difficult to draw firm conclusions
regarding the properties of the individual stars which are responsible
for the ionization of the nebulae.  It would appear safe to conclude
however, that unless the ionizing stars are highly obscured, they do
not lie at the distance of M31.  This provides another piece of
evidence for And~IV lying a significant distance beyond M31.

\subsection{Gas-Phase Metallicities}  

Optical spectra of four of the emission-line sources reveal those in
the immediate vicinity of And~IV (\#3,4,5,6) to be high-excitation  HII
regions, displaying the usual bright [OII] and [OIII] lines and, in
some cases, even marginal detections of the faint temperature-sensitive
[OIII]4363{\AA} line (see Figure \ref{specplts}).  Objects \#7,8, which
lie further away  from the centre of And~IV,  appear to be a possible
symbiotic nova and a high excitation planetary nebula respectively.
No spectra were obtained for Objects \#1 and 2.

Emission line fluxes were measured via Gaussian fits to the line
profiles.  The logarithmic extinction at H$\beta$, C(H$\beta$), was
derived from measurements of the Balmer lines, using the equation
$$\mathrm
{\frac{I_{\lambda}}{I_{H\beta}}}=\frac{F_{\lambda}}{F_{H\beta}}~10^{{C(H\beta)}{f(\lambda)}}$$
where I$_{\lambda}$ is the intrinsic line flux, F$_{\lambda}$ is the
observed line flux, and f($\lambda)$ is the Galactic reddening function
normalized to H$\beta$.  The reddening function of \citet{seat79}, was
adopted, as parametrized by  \citet{how83}, and assuming
R=A$_{V}$/E(B$-$V)=3.1.  Intrinsic case B Balmer line ratios were taken
from \citet{ost89}, assuming an electron density of N$_e$=100 cm$^{-3}$
and an electron temperature T$_e$=10$^4$~K.  The values of C(H$\beta$)
derived from the H$\alpha$/H$\beta$, H$\gamma$/H$\beta$ and
H$\delta$/H$\beta$ ratios agreed within the formal errors and no trend
was apparent to indicate the presence of Balmer absorption in the
underlying continuum.   Given its higher accuracy, we adopt the value
of C(H$\beta$) determined from the H$\alpha$/H$\beta$ ratio in our
analysis.  The derived values of the logarithmic extinction translate
into E(B$-$V)~$\simeq$~0.0$-$0.11, and indicate that the line-of-sight
extinction towards And~IV is very low and almost entirely Galactic.

Formal errors in the derived line ratios were determined by summing in
quadrature the statistical noise from the photon counts, the
uncertainty in the continuum placement (proportional to the width of
the line times the rms in the nearby continuum) and the uncertainty in
the flux calibration.  In addition, the error in C(H$\beta$) was
accounted for in deriving the reddening-corrected line ratios.  Table
\ref{linerat} presents the reddening-corrected line strengths (relative
to H$\beta$) for the three brightest HII regions as well as some
relevant line ratios. Formal errors are indicated in parentheses.
 
Oxygen and nitrogen  abundances for the HII regions were derived using
the well-established $`$semi-empirical' abundance calibrations proposed
by \citet{mcg91,mcg94} and \citet{thur96} via the procedures described
in \citet{fer98a}.  These calibrations are based on the strengths of
the bright oxygen, nitrogen and Balmer lines of hydrogen via the
parameter $$\mathrm R_{23}=\frac{\left[OII\right] \lambda
3727+\left[OIII\right] \lambda\lambda 4959,5007}{H\beta}.$$ A single
value of R$_{23}$ uniquely specifies O/H over most of the range in
metallicity, however there is a  turnover region (20--50\% solar) where
the relationship becomes double valued.  McGaugh (1994) advocates the
use of the [NII] $\lambda$6584/[OII] $\lambda$3727 ratio as a way to
discriminate between upper and lower branches, noting that it varies
monotonically with O/H and is not very sensitive to the ionization
parameter since both ions have similar ionization potentials. The
division between upper and lower branches is fairly well-defined, with
the reddening-corrected log([NII]/[OII])$> -$1 indicating the upper
branch and  log([NII]/[OII])$< -$1 indicating the lower branch.
Measurements of this line ratio in the HII regions under study here
place them all securely on the lower, metal-poor branch of the R$_{23}$
relation (see Table \ref{linerat}).  In most cases, the
[NII]$\lambda$6548 line was too faint to measure accurately so we have
assumed the theoretical value of
[NII]$\lambda$6548$=$[NII]$\lambda$6584/2.95 \citep{men82} in
calculating the N/O ratio.  The derived oxygen abundances and
nitrogen-to-oxygen ratios are low, ranging from 7$-$9\% and 9$-$16\%
the solar value respectively.  The dominant uncertainties in these
estimates are the uncertainties in the model calibrations themselves,
which are estimated to be  $\pm$0.2 dex for log(O/H) and $\pm$0.1 dex
for log(N/O) (see \citet{fer98a} for a detailed discussion). With this
in mind, there would appear to be little evidence for an intrinsic
metallicity dispersion amongst these HII regions, although Obj \#4 does
seem marginally enhanced in N/O compared to the other two objects.

How do these chemical abundances compare to those of the M31 disk at
deprojected location of And~IV?  Assuming a position angle of
35$^{\circ}$ and an inclination of 77.5$^{\circ}$ for the M31 disk, we
calculate that And~IV would have a deprojected radius of
108.4\arcmin\ or $\simeq$25~kpc if in the disk.  The chemical abundance
gradient at large radii in M31 is surprisingly poorly constrained, with
the most distant measured HII region lying at only $\sim$16~kpc
\citep{denn81}.   \citet{zar94} quote values of 12+log(O/H)$=$9.03 at
R$=$0.4R$_{25}$ and a gradient of $-$0.28 dex/R$_{25}$ for M31
(normalised to their adopted value of R$_{25}=$77.4\arcmin\ ) derived
from the measurements of \citet{denn81} and \citet{bla82}. Simple
extrapolation of this gradient predicts a gas-phase oxygen abundance of
$\sim$70\% solar at the location of And~IV.  The HII regions in the
vicinity of And~IV therefore have metallicities which are roughly an
order of magnitude lower than that expected for M31 disk gas at that
radius.  This finding adds to the mounting evidence that the emission
line sources in the vicinity of And~IV  are not associated with the
disk of  M31.

\subsection{Radial Velocities}

Our long-slit spectra also provide a measurement of the radial velocity
of each emission-line source.   The average velocity and standard
deviation determined from the observed wavelengths of the well-detected
bright H$\beta$, [OIII]5007 and H$\alpha$  lines are reported for each
HII region in Table \ref{linerat} and are in excellent agreement.
These velocities have been corrected by $\sim$15.5~km/s to account for
the motion of the Earth around the sun. The mean of these averages is
256$\pm$9~km/s, which can be compared to the value of 248$\pm$47~km/s
derived from the absorption lines in the underlying galactic continuum
(see Figure \ref{specplts}).  Both velocities differ significantly from
the radial velocity of $-$375~km/s expected for M31's disk at the
projected location of And~IV \citep{emer74}, which is also the velocity
reported previously in the literature for And~IV. Our results therefore
not only establish a direct association between the HII regions and
underlying galactic continuum emission, but also provide the final
piece of evidence that And~IV is unassociated with the disk of M31.
We adopt a heliocentric radial velocity of 256$\pm$9~km/s for And~IV.

It is tempting to speculate on whether the spread of velocities that we
measure in the And~IV HII regions can be considered real.
Interestingly, there does appear to be a systematic change of
$\sim$30~km/s in the radial velocity across the face of And~IV going
from Objects \#3 through 4 to 6, all of which were measured with a
single slit position.  Given the uncertainties in centroiding the
emission lines, as well as the rms residuals of the wavelength solution
($\sim$~0.5\AA), and the moderate resolution of our spectra, the
significance of this gradient should be considered marginal at
present.

\section{Discussion \label{disc}}

We have presented a set of new observations which constrain the nature
of the enigmatic object And~IV.   We find compelling evidence that
And~IV is \emph{not}  a star cluster lying in the disk of M31. This
evidence includes: ({\it i}) the fact that the stellar population of
And~IV does not resolve to the same extent as that of the M31 field
population, ({\it ii}) the discovery of individual HII regions in the
vicinity of And~IV which do not appear to be ionized by stars at the
distance of M31, ({\it iii}) the metallicities of these HII regions are
an order of magnitude lower than that expected for the M31 disk at the
projected location of And~IV, and ({\it iv}) the radial velocity of
And~IV differs by $\gtrsim$600~km/s from that expected for the SW side
of the M31 disk.  Furthermore, the large velocity
difference between And~IV and M31 also makes it very unlikely that
And~IV is even a bound satellite of M31. The radial velocities of the
known Andromeda satellites generally lie within $\sim$100~km/s of the
systemic velocity of M31 \citep{mat98}, whereas And~IV differs from
that by more than 500~km/s.

Two questions therefore remain: what is the true nature of And~IV and
where exactly does it lie?   We begin by addressing the second issue,
since it has bearing on the first.  Given our measured heliocentric
velocity, it is possible to use  dynamical considerations to place a
limit on the distance of And~IV. Adopting the linear Virgocentric
infall model of \citet{sch80} with parameters {$\gamma=2$},
V$_{helio}$(Virgo)=976 km~s$^{-1}$, $\omega_{\sun}$=220 km~s$^{-1}$
\citep{bing87} and D$_{virgo}$=15.9 Mpc (i.e.  H$_o$=75
km~s$^{-1}$~Mpc$^{-1}$), And~IV's position and heliocentric velocity
imply a distance of 7.0~Mpc.  Changing the heliocentric velocity of
And~IV by $\pm 90$~km~s$^{-1}$ (ie. the typical magnitude of  peculiar
motion velocities) has the effect of changing the derived distance by
$\simeq 1.2$~Mpc, ie. by $\pm 20\%$.

Next, we return to the intriguing $`$bump' seen in the  RGB LF of the
stars in the vicinity of And~IV, which we tentatively associate with
the tip of the red giant branch (TRGB) population for a distant stellar
system.  The TRGB magnitude is known to be very stable  at M$_I \simeq
-$4 over a wide range of ages (2--15~Gyr) and metallicities, and is
widely used as a distance indicator for resolved stellar systems
\citep{lee93}.  In galaxies with a significant intermediate-age
population,
 the presence of luminous asymptotic giant branch stars above the RGB
tends to smear out the edge defining the tip; this effect is very
likely to be present in And~IV and implies that the distance we derive
should be considered as a lower limit to the true distance.  From
visual inspection, we identify the TRGB from the I-band LF to lie at
I$=$25.0$\pm$0.5 and derive  a distance modulus of 29$\pm$0.5,
corresponding to a linear distance of 6.3$\pm$1.5~Mpc.  The agreement
between this distance determination and that from a dynamical argument
is extremely encouraging, especially given the significant
uncertainties in each. Taking the average of these estimates, we
therefore place And~IV at a distance of $\simeq 6.7 \pm 1.5$~Mpc.
Reassuringly,  this distance implies that the most luminous blue stars
detected in the vicinity of the And~IV  HII regions would have absolute
magnitudes in the range of M$_V \sim -5$ to $-7$, and would therefore
correspond to luminous OB stars and clusters, broadly consistent with
the observed ionization.

We now turn to clarifying the nature of And~IV.  The properties
established in this paper  -- moderately low surface brightness, very
blue colour and low metallicity -- are reminiscent of those observed in
$`$typical' low mass dwarf irregular galaxies (eg.
\citet{mat98,mill94}).   And~IV's extent of $\sim 40$\arcsec\, and disk
scalelength of $\sim 11$\arcsec\, correspond to linear sizes of
$\sim 1.3$~kpc  and 360~pc respectively at a distance of 6.7~Mpc, confirming
that the galaxy is indeed physically small.  Many dwarf irregulars
exhibit a small amount of ongoing star formation, with rates ranging
from 0.0001$-$0.01~M$_{\odot}$~yr$^{-1}$ \citep{mat98,mill94,dah93}.
Summing the \ha\ flux (corrected for Galactic reddening but not that
internal to And~IV itself) from the 5 HII regions detected within
$\simeq$1\arcmin\ of And~IV, we derive a current star formation rate of
2.6$\times~10^{-5}$~(D/Mpc)$^{2}$~M$_{\odot}$~yr$^{-1}$ using the
proportionality between SFR and H$\alpha$ luminosity derived by
\citet{kenn94}.  At our derived distance of 6.7~Mpc, this translates into
0.001~M$_{\odot}$~yr$^{-1}$ and is therefore highly consistent with the
rates measured in local dwarfs.  Yet another constraint is provided by
the gas-phase metallicity of And~IV.  Based on the mean oxygen
abundance of $\sim 10\%$~solar measured for the And~IV HII regions, the
metallicity-luminosity relation of \citet{skill89} predicts M$_B=-$15
which, once again, supports the identification of And~IV as a dwarf
galaxy.   
 
In terms of properties such as star formation rate, metallicity,
central surface brightness and inferred luminosity, And~IV appears very
similar to Local Group dwarf irregulars IC~1613 and Sextans~A
\citep{mat98}.  High-quality HST CMDs have recently been published for
both of these systems \citep{dp97,cole99}, and we consider how these
diagrams may help us to better understand the nature of the $`$excess'
resolved stars seen towards And~IV (Section 3.1.1).  One of the most
striking features of the Sextans~A CMD is the population of massive
core helium-burning stars (the so-called $`$blue loop' stars) seen just
redward of the main sequence; this feature is also seen, albeit to a
slightly lesser degree, in IC~1613.  As the prominence of blue loop
stars is greatest at low metallicites, and as (at least) the gas-phase
metallicity of And~IV and Sextans~A/IC~1613 are all similarly low, it
is reasonable to expect that such stars are also present in the And~IV
CMD.   The $`$excess' stars seen lying between the M31 blue plume and
RGB on the And~IV CMD, but not on that of the M31 field,  could be 
the signature of this component.   Furthermore, IC~1613 shows a
sizeable population of red supergiants extending to I$\approx$16.5 (or
V$\approx$18), which corresponds to M$_V \sim -6.5$ for a distance
modulus of 24.27 \citep{cole99}.  Given that the  $`$excess' luminous
red stars identified on the And~IV CMD would have absolute
magnitudes in the range $-7.5$ to $-6.5$ at our derived distance, it appears likely that these stars are red supergiants belonging to And~IV.

An interesting question is whether And~IV is an isolated dwarf galaxy or
whether it belongs to some larger group environment. There are no
catalogued galaxy groups in the vicinity of And~IV but  a search with
NED reveals 14 galaxies in the range 22$^{h}<\alpha<0^{h}$,
20$^{\circ}<\delta<60^{\circ}$ and 150~km/s$<$V$_{helio}<$350~km/s.
Of these, the most luminous are IC~1727, NGC~784 and UGC~64. We suspect
that And~IV may be one of the many low-luminosity galaxies inhabiting
this environment.

\section{Conclusions}

We have presented deep HST WFPC2 and ground-based observations of the
enigmatic object And~IV.  The true nature of this object -- old $`$star
cloud' in the outer disk of M31 or background galaxy -- has remained a
mystery since it was first discovered by \citet{vdb72} during his
search for dwarf spheroidal companions to M31.  From the analysis of
our WFPC2 images and complementary \ha imaging and long-slit optical
spectroscopy, we find compelling evidence that And~IV is a background
galaxy seen through the disk of M31.   The moderate surface brightness
($\bar{\mu_V}\sim$24), very blue colour (V$-$I$\sim$0.6), low current
star formation rate ($\sim$0.001~M$_{\odot}$~yr$^{-1}$) and low
metallicity ($\sim$10\% solar) reported here are consistent with And~IV
being a  dwarf irregular galaxy, perhaps a more distant analog of
Local Group members IC~1613 and Sextans~A.  Indeed, such objects are
very common in the nearby Universe, and perhaps it is not surprising to
find one projected behind M31.   The distance to And~IV is
not tightly constrained by our current dataset, but arguments based on
both the observed radial velocity and on a tentative detection of the
RGB tip suggest it lies in the range 5$\lesssim$D$\lesssim$8~Mpc,
placing it well outside the confines of the Local Group.  At this
distance, the physical extent of And~IV is consistent for what is
expected of small dwarf galaxies.  And~IV may
belong to a loose, previously uncatalogued group, containing major
members UGC~64, IC1727 and NGC~784.

\acknowledgments

We are grateful to Rachel Johnson, Mike Irwin and Nial Tanvir for
useful advice and discussions. Gerhardt Meurer is thanked for
assistance with Virgocentric infall calculation and Piero Rosati for help
with the radial velocity cross-correlation.  This research has made use
of the NASA/IPAC Extragalactic Database (NED) which is operated by the
Jet Propulsion Laboratory, California Institute of Technology, under
contract with NASA. Support for this work was provided by NASA through
grant number GO-067340195A from the Space Telescope Science Institute,
which is operated by AURA, Inc. under NASA contract NAS5-26555.\\

\clearpage

\begin{figure}
\figurenum{1a}
\caption[figure1.ps]{An image of M31
constructed from scanned DSS red and blue plates (courtesy Richard
Sword (IoA)).  The small box indicates the region that is enlarged in
Figure 1(b).}
\vspace*{2cm}
\figurenum{1b}
\caption{ An image constructed from V and I data taken with
the WIYN~3.5m telescope.  The data was taken under excellent
seeing conditions (0.6$-$0.7$''$) and the FOV is $\simeq$6\arcmin.
And~IV reveals itself as fuzzy (blue) blob extending westwards from the
central star.  The approximate area covered by the WF3 chip is indicated on the
image.  North is to the top and east to the left.  The WIYN Observatory
is a joint facility of the University of Wisconsin-Madison, Indiana
University, Yale University, and the National Optical Astronomy
Observatories.}
\vspace*{3cm}
\figurenum{2}
\caption[figure2.ps]{Colour image constructed from the HST WF3 F555W
and F814W images. And~IV is visible in the lower left of the chip.  
\label{wfima}}
\figurenum{3}
\vspace*{3cm}
\caption[figure3.ps]{(a) \ha image of And~IV and the surrounding
field taken with the INT 2.5m. (b) \ha continuum-subtracted version
of this image.  North is to the top, and east to the left. The size of
each image is $\sim 8\times8$\arcmin.  Emission line sources are labelled
and the WFPC2 FOV is overlaid. \label{haima}}
\end{figure}

\begin{figure}
\figurenum{4}
\centerline{\psfig{file=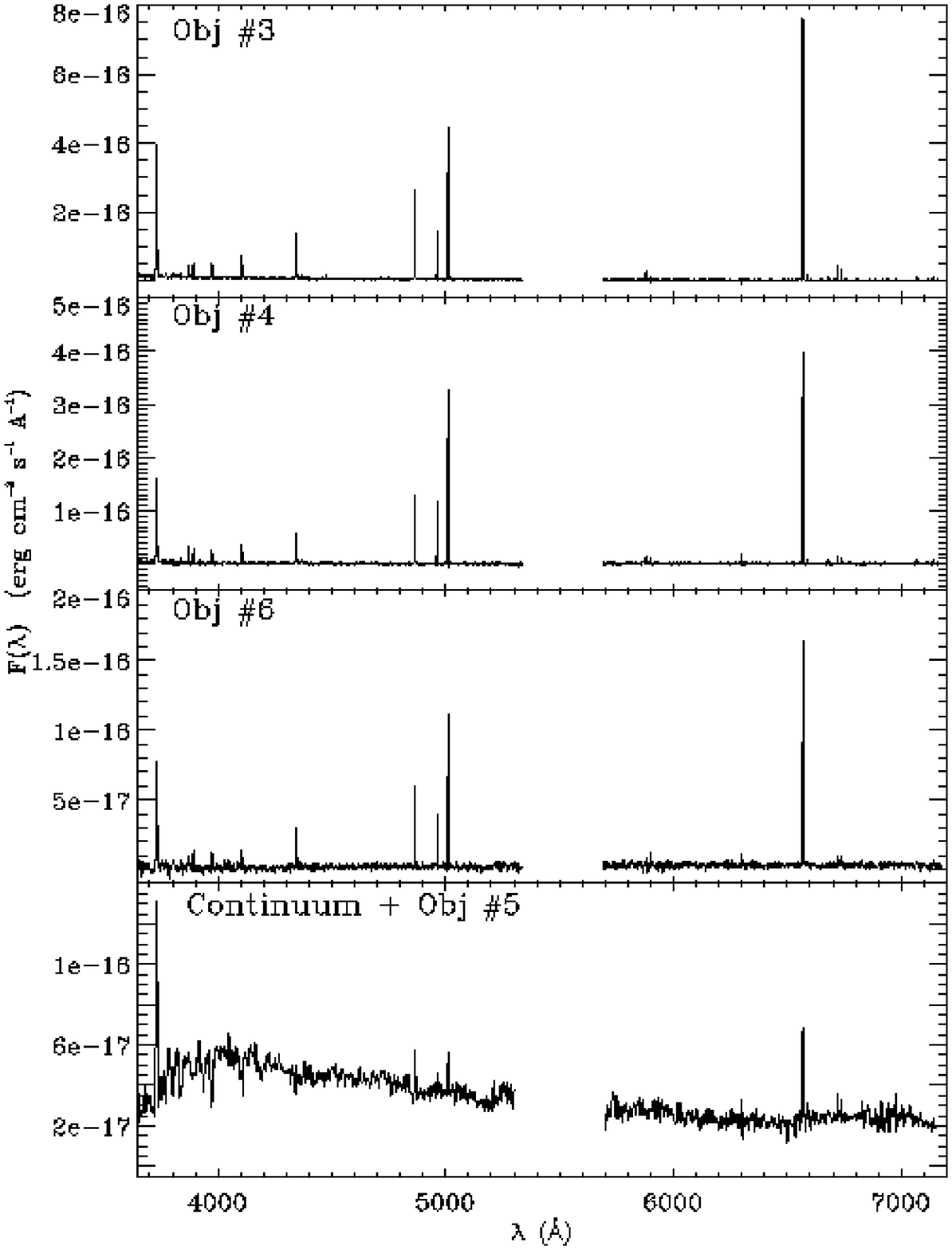,height=15cm}}
\caption[figure4.ps]{Spectra taken with the WHT 4.2m for the three
brightest nebulae  in the vicinity of And~IV and for a $\sim$20\arcsec\ 
region of the underlying
galactic continuum emission (bottom). This aperture also includes
Obj \#5.  \label{specplts}}
\end{figure}

\begin{figure}
\figurenum{5}
\centerline{\psfig{file=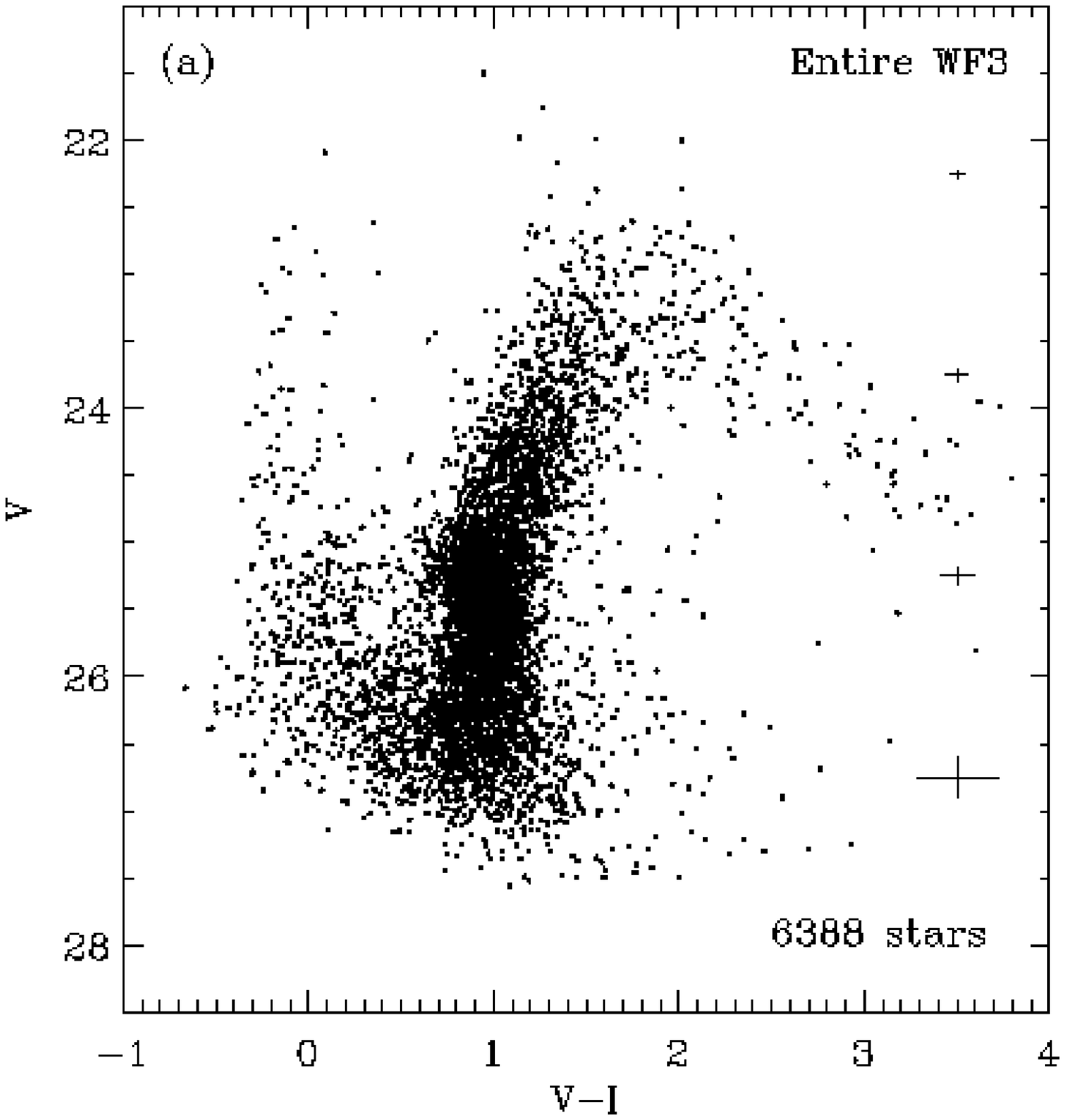,width=9cm}}
\vspace*{.5cm}
\centerline{\psfig{file=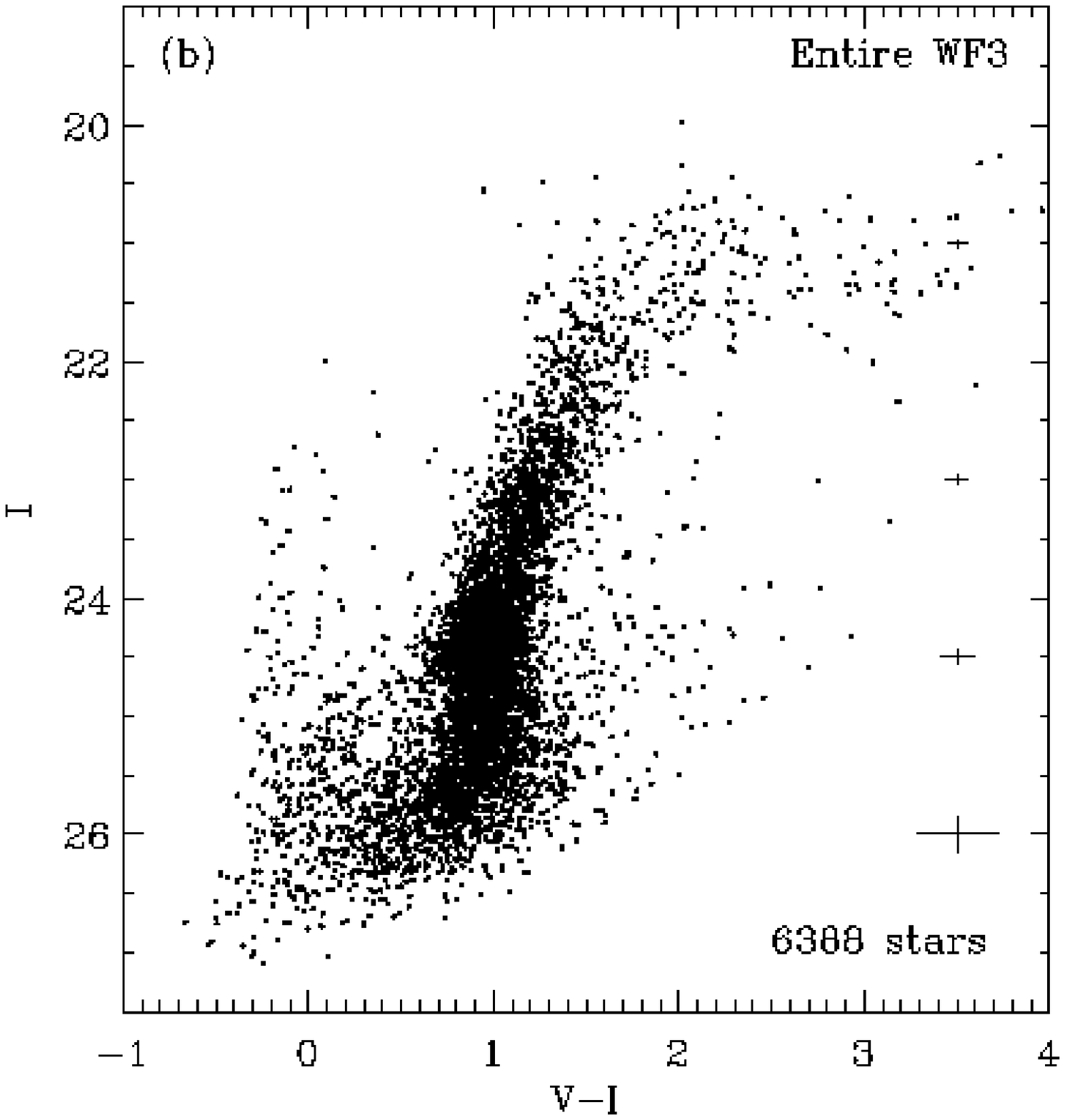,width=9cm}}
\caption[figure5.ps]{(a) (V, V$-$I) colour-magnitude-diagram for
6388 stars detected on the entire WF3 chip. Mean photometric errors are
indicated on the right.   (b) (I, V$-$I) colour-magnitude-diagram
for the same stars. \label{cmds}}
\end{figure}

\begin{figure}
\figurenum{6}
\centerline{\psfig{file=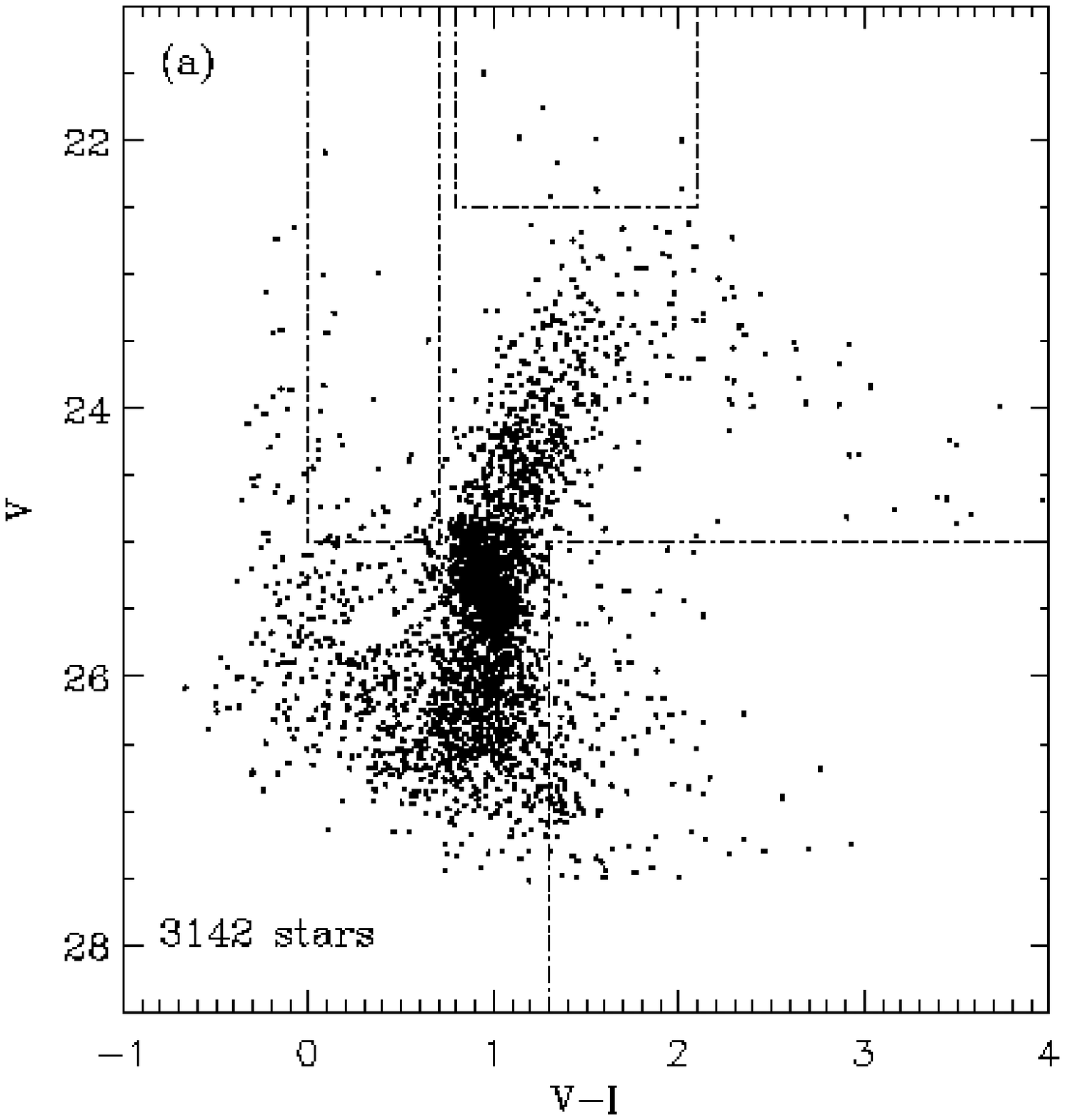,width=9cm}}
\vspace*{.5cm}
\centerline{\psfig{file=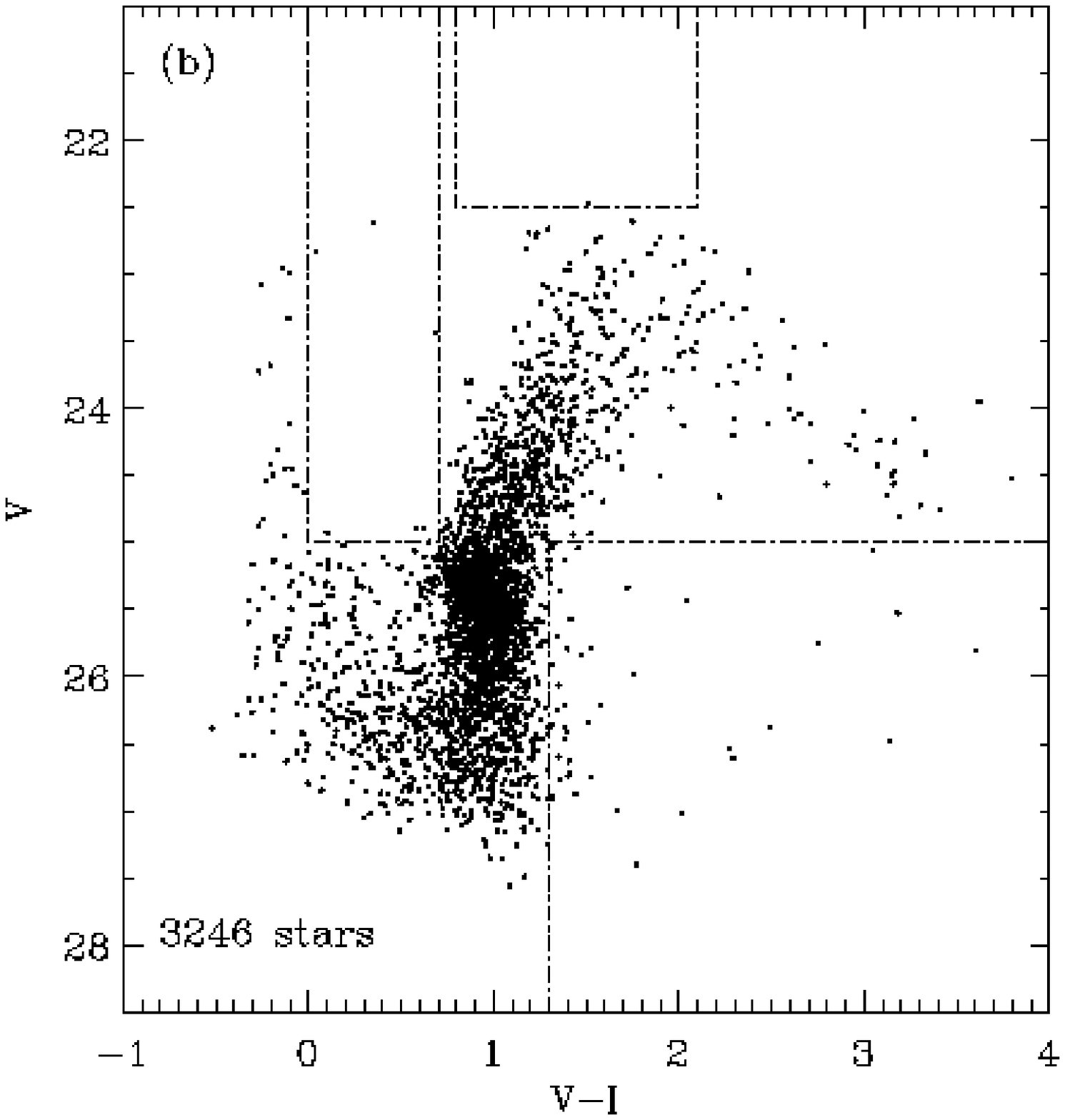,width=9cm}}
\caption[figure6.ps]{(V, V$-$I) CMD for stars detected within a box of
40$''\times$60$''$ centered on And~IV (a) and for stars lying on the
remaining area of WF3, ie. the M31 field (b).  The boxes indicate portions of the CMDs that appear to be significantly more populated on the $`$And~IV'
CMD than on the $`$off' one.   \label{compcmds}}
\end{figure}

\begin{figure}
\figurenum{7}
\centerline{\psfig{file=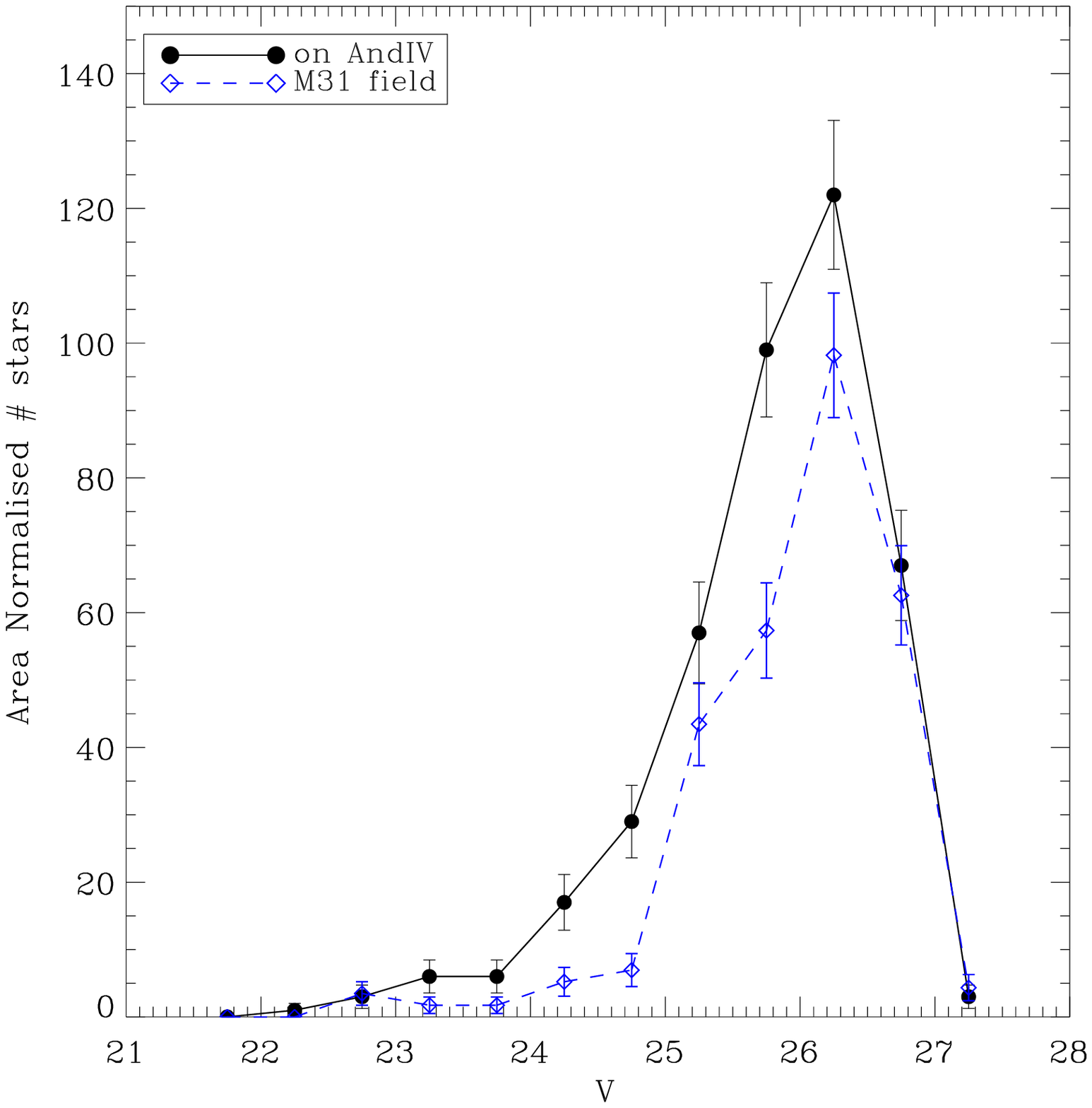,width=9cm}}
\vspace*{.5cm}
\centerline{\psfig{file=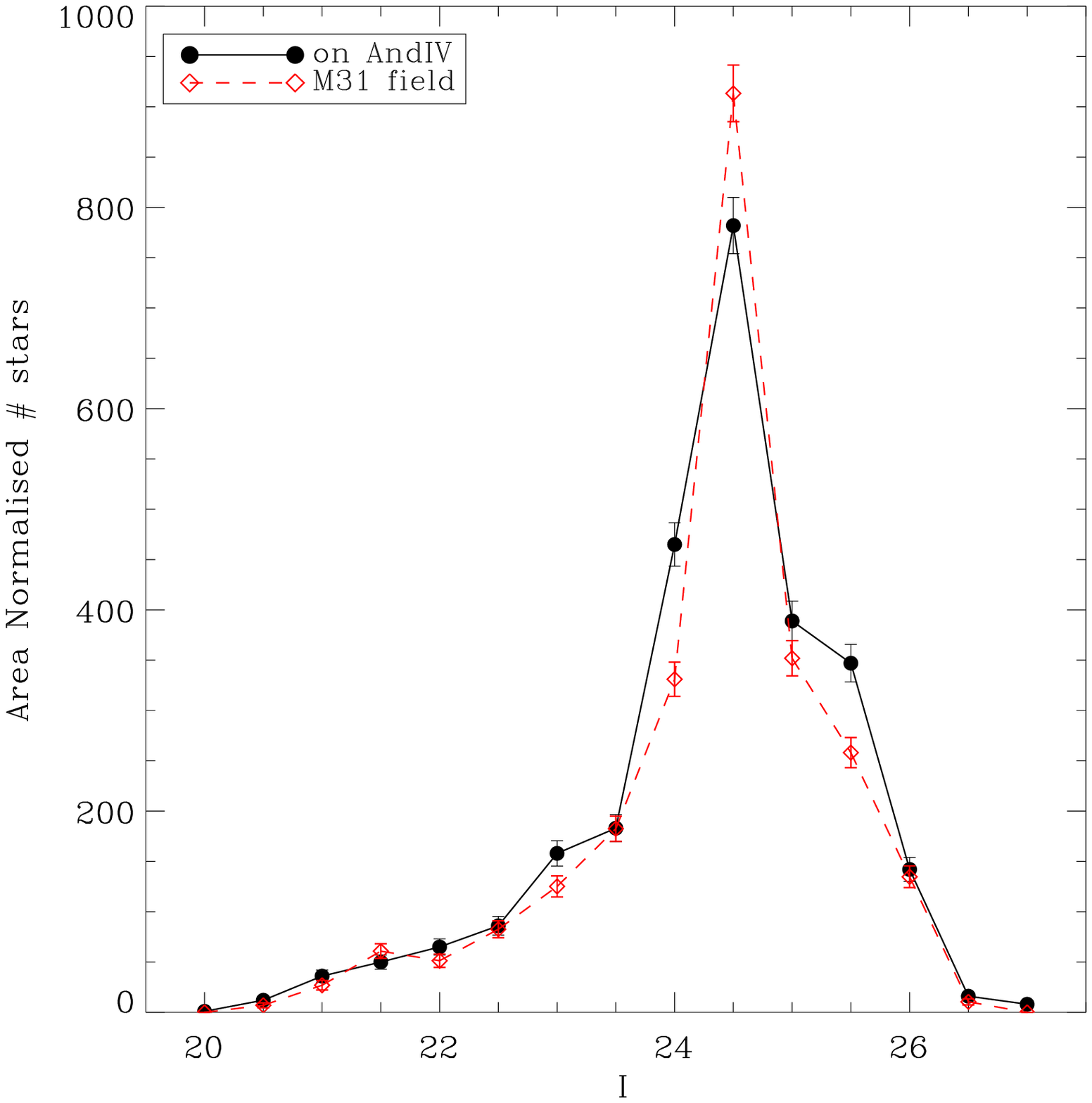,width=9cm}}
\caption[figure7.ps]{ (a) Main-sequence V-band LFs for stars
on and off And IV. Main-sequence stars are defined as all stars lying 
blueward of V$-$I$>$0.6. (b) RGB I-band LFs constructed for stars on
and off AndIV.  RGB stars are defined here as all stars lying 
redward of V$-$I$>$0.6.  A correction has been made for the slightly
different areas occupied by these regions. Poissonian error
bars are indicated.  \label{lumfs}}
\end{figure}

\begin{figure}
\figurenum{8}
\vspace*{15cm}
\caption[figure8.ps]{The boxcar-smoothed residual F555W image.  The
cross indicates the adopted centre of And~IV. \label{resima}}
\end{figure}

\begin{figure}
\figurenum{9}
\centerline{\psfig{file=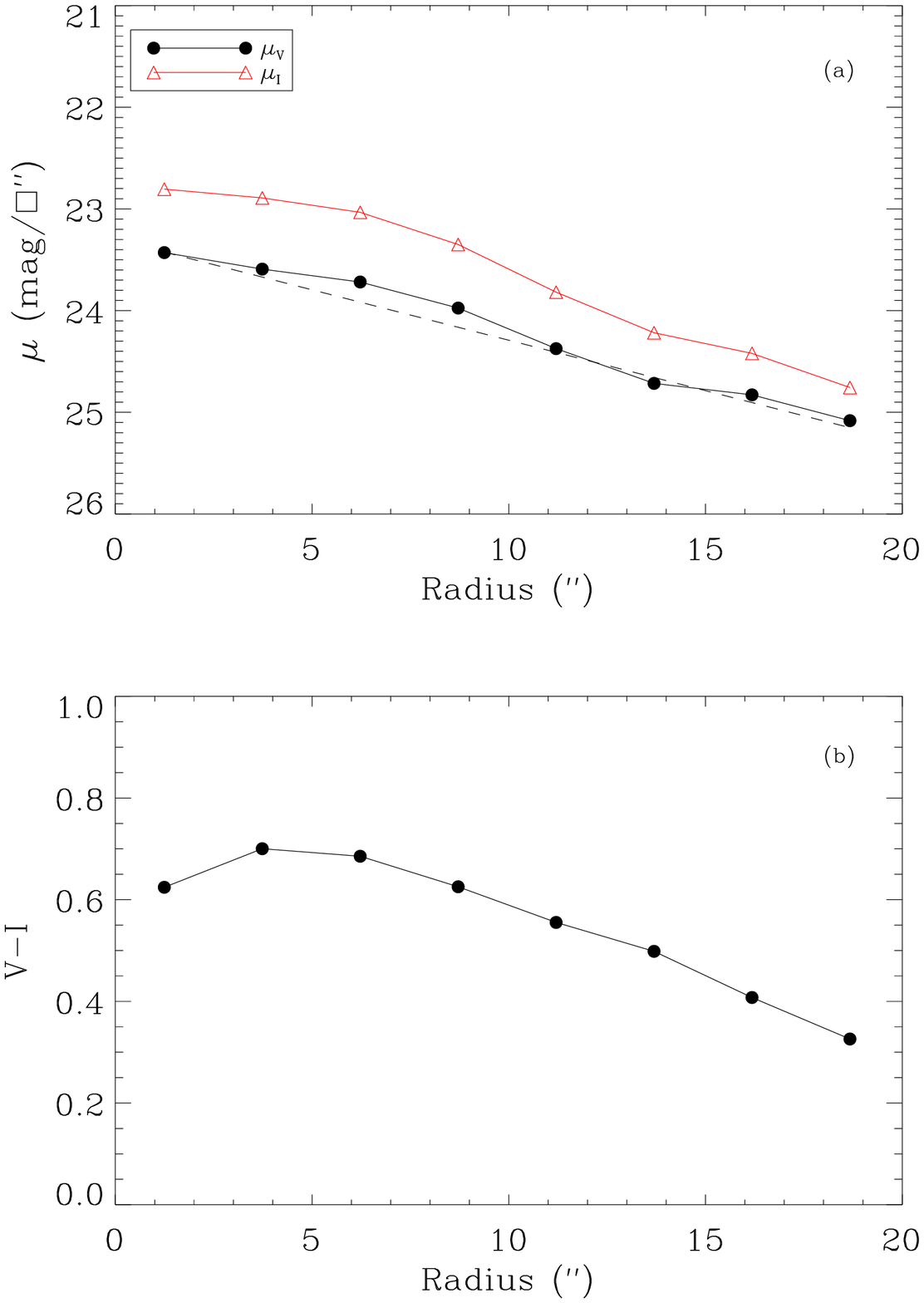,height=18cm}}
\caption[figure9.ps]{(Top) V- and I-band surface brightness
profiles for the residual emission around And~IV, derived from elliptical aperture photometry.  
The profile of an exponential disk with $\mu_{V}=$23.3 and
$\alpha_{V}=$11\arcsec\ is overlaid for comparison.  (Bottom) V$-$I 
colour gradient as a function of radius.    \label{sbprofs}}
\end{figure}

\newpage

\begin{deluxetable}{cccccc}
\tablecaption{Properties of Emission Line Sources in the Vicinity of And~IV. \label{emsrcs}}
\tablewidth{0pt}
\tablehead{
\colhead{ID} & \colhead{RA (2000)}   & \colhead{DEC (2000)}   &
\colhead{F(H$\alpha$)\tablenotemark{b}} & \colhead{R$_{proj}$\tablenotemark{c}} & 
\colhead{Comments\tablenotemark{d}}  
}
\startdata
1 & 00:42:37.6 & 40:38:12 & 3.58$\times$10$^{-15}$ & 241 &u,c \\
2 & 00:42:36.6 & 40:34:04 & 1.61$\times$10$^{-15}$ & 51&r,c\\
3 & 00:42:32.3 & 40:33:57 & 9.26$\times$10$^{-15}$ & 22 & r,c\\
4 & 00:42:31.8 & 40:34:10 & 2.95$\times$10$^{-15}$ & 11&r\\
5 & 00:42:32.8 & 40:34:16 & 1.28$\times$10$^{-15}$ & 6&r,c?\\
6 & 00:42:30.6 & 40:34:46 & 2.11$\times$10$^{-15}$ & 34&r\\
7 & 00:42:20.3 & 40:32:30 & 2.71$\times$10$^{-15}$ & 175&u\\
8 & 00:42:21.2 & 40:33:48 & 3.78$\times$10$^{-15}$ & 131&u\\
\enddata


\tablenotetext{a} {Coordinates are accurate to $\pm$1\arcsec.}
\tablenotetext{b}{F(H$\alpha$) in erg~s$^{-1}$~cm$^{-2}$. Reported values are corrected
for [NII] contamination and Galactic extinction.}
\tablenotetext{c} {Projected distance (measured in \arcsec) from the centre of And~IV.}
\tablenotetext{d} {Comments refer to whether the source is resolved (r) or unresolved (u) on our images, and to whether it has a continuum counterpart
(c).}

\end{deluxetable}

\begin{deluxetable}{cccc}
\tablecolumns{7}
\tablewidth{0pc}
\tablecaption{\label{linerat} Line Intensities for HII Regions in the Vicinity of And~IV}
\tablehead{\colhead{Line} & \colhead{Obj \#3} & \colhead{Obj \#4} & \colhead{Obj \# 6}}
\startdata
$[OII]$ 3727 &   2.56(0.14) &  1.81(0.10) & 1.88(0.12) \\
H$\delta$ 4101 & 0.31(0.02) &  0.30(0.02) & 0.20(0.02) \\
H$\gamma$ 4340 & 0.50(0.02) & 0.49(0.03) & 0.51(0.03) \\
H$\beta$ 4861 & 1.00(0.06) & 1.00(0.06) & 1.00(0.07) \\
$[OIII]$ 4959 & 0.60(0.03) & 1.06(0.06) & 0.57(0.04) \\
$[OIII]$ 5007 & 1.94(0.11) & 3.05(0.18) & 1.96(0.14)\\
H$\alpha$ 6563 & 2.86(0.18) &  2.86(0.18) & 2.58(0.19)\\
$[NII]$ 6583 &  0.09(0.01) & 0.08(0.01) & 0.06(0.01)\\
$[SII]$ 6717 & 0.13(0.01) & 0.13(0.01) & 0.14(0.02)\\
$[SII]$ 6731 & 0.09(0.01) & 0.09(0.01) & 0.10(0.02)\\
\tableline
EW(H$\beta$)({\AA}) & 171. &  684. & 148.3 \\
F(H$\beta)$ (erg~s$^{-1}$~cm$^{-2}$) & $8.8\times10^{-16}$ &  $4.5\times10^{-16}$ & $2.3\times10^{-16}$ \\
C(H$\beta)$ & 0.16(0.06) & 0.0(0.06) & 0.0(0.07)\\
log($[NII]6584/[OII]$) & -1.47(0.02) &  -1.34(0.02) & -1.53(0.03)\\
log($[OII]/[OIII]$) & 0.00(0.03) &  -0.36(0.03) & -0.13(0.04)\\
log(R$_{23}$)& 0.71(0.02) & 0.77(0.02) & 0.65(0.02) \\
log(O/H) & -4.10 & -4.12 & -4.25 \\
log(N/O) &  -1.83 & -1.66  & -1.93 \\
V$_{helio}$ (km/s)&  244(15) & 250(13) & 273(19) \\
\enddata

\end{deluxetable}

\end{document}